\documentclass[aps,amsmath,amssymb,reprint,groupedaddress,floatfix]{revtex4-1}

\usepackage{graphicx}
\usepackage{bm}

\usepackage{amsmath}
\usepackage{amssymb}
\usepackage{color}
\usepackage{times}
\usepackage{float}
\usepackage[sort&compress]{natbib}

\usepackage[colorinlistoftodos,prependcaption,textsize=small]{todonotes}

\usepackage{hyperref}

\hypersetup{
  colorlinks=true,
citecolor=black,
linkcolor=black,
urlcolor=black
  }

\usepackage[T1]{fontenc}
\usepackage[utf8]{inputenc}

\begin{document}

\title{A Stochastic Model of Translocation of Knotted Proteins}
\author{Karol Capa{\l}a}
\email{k.capala@sanoscience.org}

\affiliation{Personal Health Data Science Group, 
Sano - Centre for Computational Personalised Medicine}
\affiliation{Institute of Theoretical Physics, Department of Statistical Physics, Jagiellonian University}

\author{Piotr Szymczak}
\email{piotrek@fuw.edu.pl}
\affiliation{Institute of Theoretical Physics, Faculty of Physics, University of Warsaw}



\begin{abstract}
Knotted proteins, when forced through the pores, can get stuck if the knots in their backbone tighten under force. Alternatively, the knot can slide off the chain, making translocation possible. We construct a simple energy landscape model of this process with a time-periodic potential that mimics the action of a molecular motor. We calculate the translocation time as a function of the period of the pulling force, discuss the asymptotic limits and biological relevance of the results.
\end{abstract}

\maketitle

\setlength{\tabcolsep}{0pt}

%
%
\section{Introduction}
It is increasingly realized that topology plays an important role for the functional and dynamical properties of biomolecules. Knots and tangles are rather unavoidable in the DNA chain, due to its huge length and large density, with 2 meters of DNA squeezed in a tiny cell nucleus. However, topology can also be important for function, dynamics and stability of proteins, although the exact function of the entanglement in proteins is still under debate~\cite{faisca2015,perego2019,jackson2020,sulkowska2020}. In about
1\% of the proteins the polypeptide chain~\cite{Virnau2006,jamroz2015}, which forms a protein backbone, adopts a knotted configuration. It was reported that the presence of knots can increase the thermal and mechanical stability of proteins~\cite{Sulkowska2008stabilizing,dabrowski2016} or help them in their enzymatic activity~\cite{jacobs2002,christian2016,Ko2019}, but it can also be hindering, particularly during folding~\cite{King2010,Mallam2012,rivera2020}, unfolding~\cite{Ziegler2016}, and passing through narrow constrictions~\cite{szymczak2013,san2017}.  The latter happens when the protein is degraded in proteasome or translocated through the intercellular membranes, e.g. during import into mitochondria~\cite{Pfanner1995,Matouschek2000,Mokranjac2005,Panja2013,muthukumar2014}. The unfolding and import of proteins into mitochondria or proteasome are facilitated by molecular motors that act with forces of the order of 30pN~\cite{Alder2003}. However, as shown in a number of studies, both experimental and numerical ~\cite{Sulkowska2008tightening,Bornschlogl2009,Dzubiella2009}, the protein knots tend to tighten under the action of the force. The radius of gyration of the tight knot has been estimated to be around 7–8{\AA} for the simplest protein knot (a trefoil) and correspondingly larger for more complicated knots. On the other hand, the smallest constrictions in mitochondrial pores or proteasome openings are 6–7{\AA} in radius \cite{Forster2003,Rehling2004}, thus proteins with knotted backbones might have problems navigating them. 

\begin{figure}[!h]
    \centering
    \includegraphics[width=0.95\columnwidth]{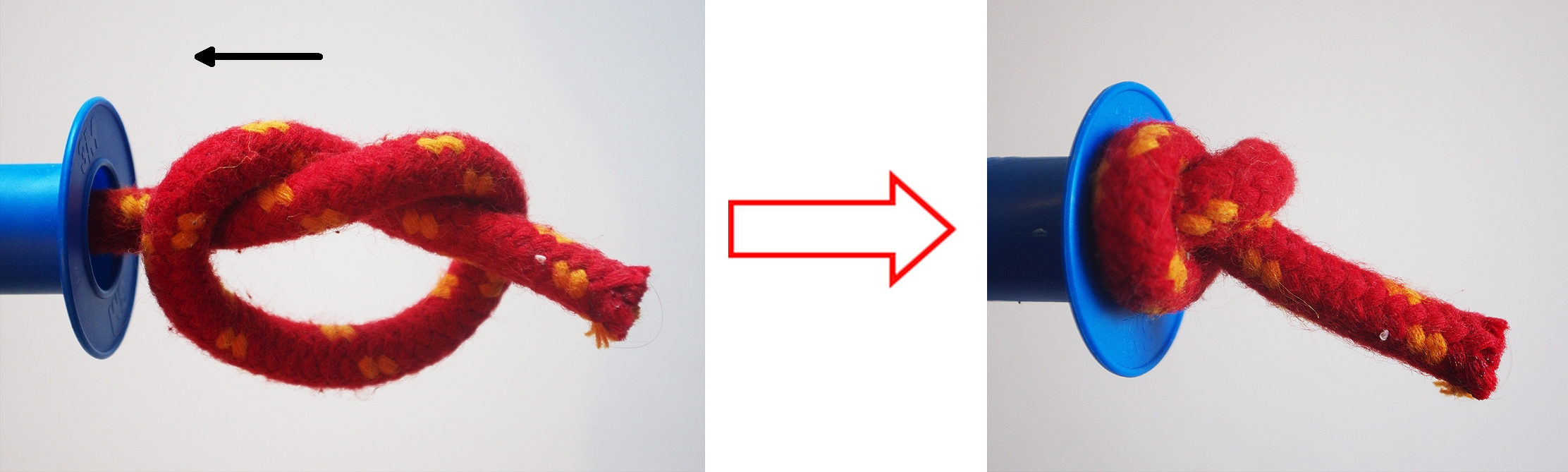}
    \caption{Tightening of the knot on a rope while pushing it through an opening with the diameter smaller than that of a tightened knot}
    \label{jamming}
\end{figure}

Assuming that the translocation of a knot is impossible, from the macroscopic point of view two possibilities remain - either the knot would slide off the rope or it would block the opening. Sewing afficionados know that very well, as they make thread knots using a narrow space between the fingers through which the thread goes easily, but the knot jams and tightens. (Fig.~\ref{jamming}). However, proteins are not like ropes or sewing threads in one important respect. Different parts of the protein chain attract each other, and as a result, the molecule - if left on its own, with no forces acting on it - folds into its native conformation (a specific spatial configuration which is critical to its biological function). 

As proposed in Ref.~\cite{szymczak2016}, such a folding propensity of proteins, together with the repetitive nature of the forces produced by molecular motors, enables the knotted molecules to translocate successfully. The molecular motors work in an on-off manner, during an "on" part of the cycle, they attempt to pull the knotted protein into the pore. During pulling, the knot slides towards the free end of the chain \cite{wang2013,san2017}. If it succeeds in sliding off the chain before it tightens, the protein translocates successfully. The tightened knot, on the other hand, jams the pore, but not permanently. During the next off-cycle of the force, as the protein begins to refold, some stored length is inserted into the knotted core, and the knot loosens, thus escaping the tightened configuration. Subsequently, during the next force-on
period the protein makes another attempt at the translocation, with an eventual success after sufficiently
many attempts. 

Ref.~\cite{szymczak2016} looked at this system through a series of molecular dynamics simulations that resolved the interactions between individual aminoacids as well as between aminoacids and the pore walls. It seems, however, that the main physical mechanism behind this process can be captured in a much more simple one-dimensional model in the spirit of reaction rate theory. The construction of such a description is the objective of the present manuscript.

\section{Model}

We will model the dynamics of our system in terms of reaction rate theory \cite{hanggi1990} with a one-dimensional reaction coordinate $x$, indicating the state of the system. Within this model, the dynamics of the system along the reaction coordinate is assumed to follow the overdamped Langevin equation~\cite{gardiner2009}
\begin{equation}
  \frac{dx}{dt}=-\mu \frac{dV(x,t)}{dx}+\sqrt{\mu kT} \xi(t),
  \label{eq:langevin}
\end{equation}
where $\xi(t)$ is the Gaussian white noise satisfying $\langle \xi(t) \rangle=0$ with $\langle \xi(t)\xi(s) \rangle =  \delta(t-s)$, $\mu$ is the effective mobility along the reaction coordinate and $T$ is the temperature.

The initial position of the knotted protein in front of the pore corresponds to the point $x=0$ in our 1d reaction model. At this point, the protein is in its native conformation - its backbone is knotted, but the knot is not tightened. In the following, we will denote this state as KL ({\it knotted loose}). The point $x=L_1$, on the other hand, represents the protein with a fully tightened knot (KT - {\it knotted tightened}), which is blocking the pore. To represent this blockage, we place an impermeable (reflecting) wall at $x=L_1$. 
Finally, the point $x=-L_2$ represents the protein which has successfully translocated through the pore and left the system. Since in this process the knot slides off the chain, such a state will be called U ({\it unknotted}) in subsequent considerations. We assume that translocation is only possible under the action of the force, so we make the wall at $x=-L_2$ absorbing only in the first part of the period (when the force is on).  In the second part of the force cycle this wall will be reflecting.

\begin{figure}[!h]
    \centering
    \includegraphics[width=0.95\columnwidth]{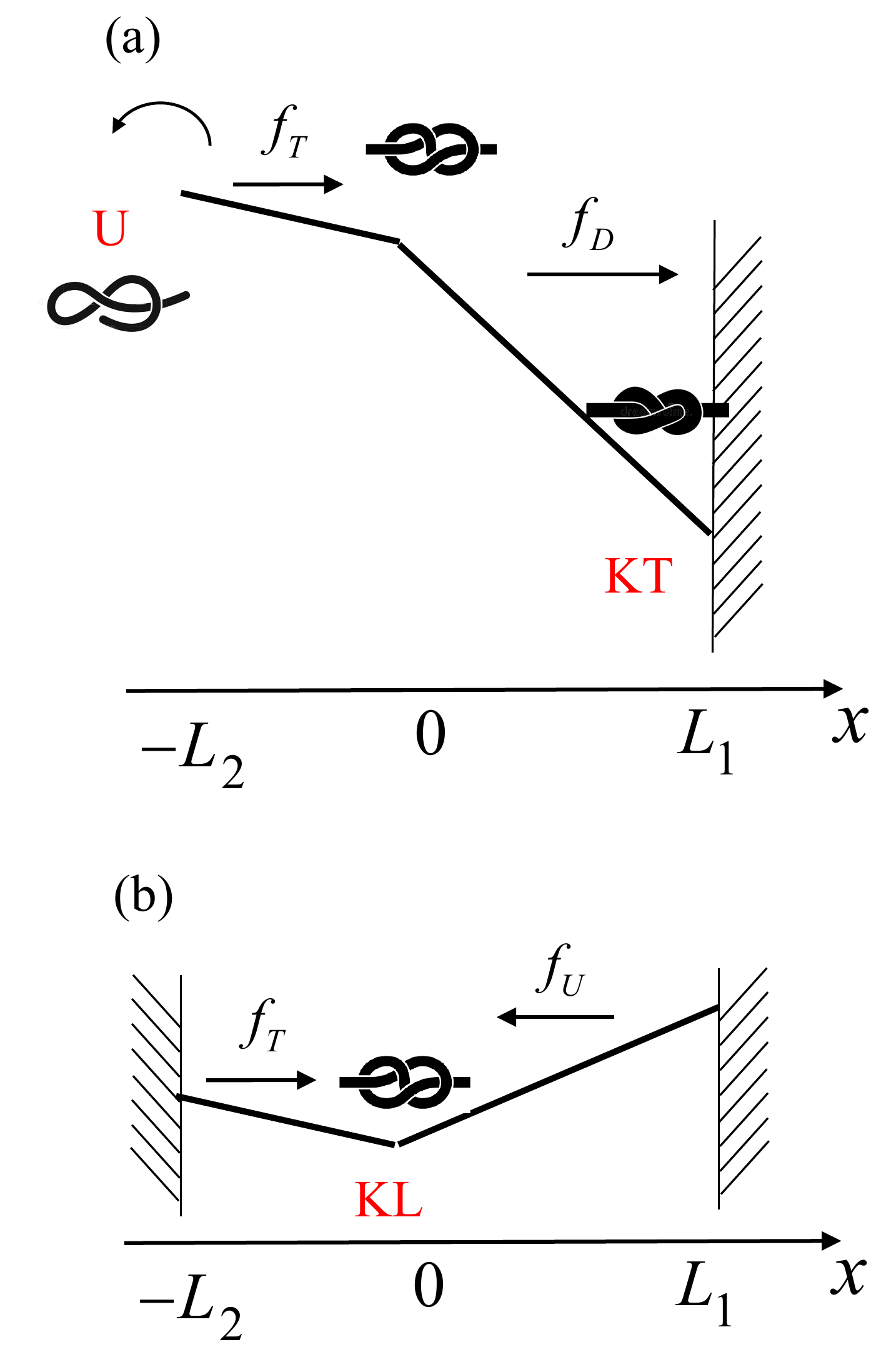}
    \caption{Potential given by Eq.~(\ref{eq:potential}) under pulling (a) and in the free state (b). The marked conformations correspond to the state with a loose knot (KL), tightened knot (KT) and untied knot (U).}
    \label{fig:potential}
\end{figure}

To mimic the repetitive nature of molecular motors, we will introduce a potential   $V(x,t)$ which switches between two states with period $2T$. To describe it, we find it convenient to introduce an additional variable.
\begin{equation}
    \mathcal{T} = \mod(t,2T),
    \label{eq:czasOdPoczatkuOkresu}
\end{equation}
which measures the time from the beginning of the present force period.
Using this variable, the periodic potential is given by
\begin{equation}
    V(x,t)= \left\{ 
    \begin{array}{l}
V_{\mathrm{pull}}(x) \quad \mathcal{T}(t) < T\\
 \\
V_{\mathrm{free}}(x) \quad \mathcal{T}(t)> T
\end{array}
\right.
\label{eq:potential}
\end{equation}
where $V_{\mathrm{pull}}(x)$ and $V_{\mathrm{free}}(x)$ are time-independent, piecewise linear potential wells (see Fig.~\ref{fig:potential}), with
\begin{equation}
    V_{\mathrm{pull}}(x)= \left\{ 
    \begin{array}{l}
 -f_T x \quad x < 0\\
 \\
  -f_D x  \quad x \geqslant 0
\end{array}
\right..
\end{equation}
and
\begin{equation}
    V_{\mathrm{free}}(x)= \left\{ 
    \begin{array}{l}
 -f_T x \quad x < 0\\
 \\
  f_U x  \quad x \geqslant 0.
\end{array}
\right.
\label{eq:potentialUp}
\end{equation}
In the first half of the force period $\mathcal{T}<T$, the potential $V_{\mathrm{pull}}(x)$ corresponds to the situation when the molecular motor exerts the force on a protein chain, pulling it in. 
The appearance of this force is represented by a potential well of depth $f_D L_1$ at $x=L_1$. As mentioned above, the trapping of the particle at this position corresponds to tightening of the knot under the action of the force. The other possible pathway corresponds to a successful translocation through $x=-L_2$ wall. 
Getting there is much less likely than entering the kinetic trap at $x=L_1$. To represent that, we assume that there is an energy barrier ($f_T L_2$) to overcome to reach the absorbing wall at $x=-L_2$.

The second part of the force period, when the force is off, is described by the potential $V_{\mathrm{free}}(x)$. The minimum at $x=L_1$ disappears and is replaced by a maximum of height $f_UL_1$. This creates a potential ramp pulling the particle towards the origin, which represents the effective action of the interactions between the aminoacids, which are trying to refold the protein into its native conformation. The other part of the potential (for $x<0$) is the same as in the first part of the force period, with the only difference that now the boundary at $x=-L_2$ is assumed to be reflecting. This represents the fact that it is impossible to translocate through the pore in the absence of force.

Before proceeding, let us introduce the dimensionless variables
\begin{equation}
\left\{
\begin{array}{lcl}
\tilde{x} & = &  x/L_1, \\
\tilde{t} & = & t \mu k T /L_1^2
\end{array}
\right..
\label{eq:transformation}
\end{equation}
and rescaled parameters,
\begin{equation}
\left\{
\begin{array}{lcl}
\tilde{f_i} & = & f_iL_1/kT,\\
\tilde{L} & = &  L_2/L_1
\end{array}
\right..
\label{eq:transformation2}
\end{equation}
After such a transformation, the motion is restricted to the $(-\tilde{L},1)$ interval, with the forces rescaled by the characteristic thermal forces.
For convenience, from now on, we will drop tildes and consequently use dimensionless units.

\section{Asymptotic regimes}

To characterize the escape dynamics of the system, we analyze the distribution of the first exit times.
The latter is defined as
\begin{equation}
    \langle \tau \rangle =
     \langle \min\{t : x(0)=0 \;\land\; x(t) \leqslant -L  \} \rangle.
\end{equation}

Due to the time dependence of the potential $V(x,t)$ the general solution is difficult to obtain.
Nevertheless, one may obtain a number of useful results for the limit of slowly-changing potential, i.e. $T\to \infty$.
In this limit, the first escape time is dominated by waiting for the state when escape is possible.
Therefore, MFET is proportional to the average number of force periods before the particle escapes
\begin{equation}
    \langle \tau \rangle = 2 \langle n \rangle T.
    \label{eq:assimptitics}
\end{equation}
To calculate $\langle n \rangle$ we assume that $f_D$ is sufficiently deep so that the force pulling the particle towards the KT state for $x>0$ is so high that the escape chance is negligible and we inevitably end up in the trap at $x=1$. In this case
\begin{equation}
    \frac{1}{\langle n \rangle} = \int_{-L}^0 p(x) \pi_L(x) dx,
    \label{eq:escapeProb}
\end{equation}
where $p(x)$ is the probability density of finding the particle in $x$ at $\mathcal{T}=0$ and $\pi_L(x)$ is the probability that the particle escapes through the left boundary, assuming that it started at $x$.

The general formula for the probability that the particle starting in the $(-L,0)$ interval escapes through the left barrier is given by\cite{gardiner2009}.
\begin{equation}
    \pi_L(x)=\frac{\int_x^0 \psi(y)dy}{\int_{-L}^0 \psi(y)dy},
    \label{eq:splitting}
\end{equation}
where, for potential (\ref{eq:potential}) 
\begin{equation}
    \psi(x)=\exp{\left( \int_{-L}^x f_T dy \right)}=\exp{\left( f_T (x+L) \right)}.
\end{equation}
In the limit of $T\to \infty$ the particle probability density will have sufficient time during the off-force period to relax towards the stationary distribution, $p_{st}(x)$. 
For the potential of the form $V_{\mathrm{free}}(x)$ the stationary distribution is
\begin{equation}
    p_{st}(x)= \left\{ 
    \begin{array}{l}
 A \exp{\left(f_T x \right)} \quad x < 0\\
 \\
  A \exp{\left(-f_U x \right)} \quad x \geqslant 0
\end{array}
\right.,
\label{eq:statState}
\end{equation}
with a normalization factor
\begin{equation}
    A=\frac{f_U}{\frac{f_U-f_U e^{-f_T L}}{f_T}+\sinh (f_U)-\cosh (f_U)+1}.
\end{equation}

On the other hand, during the on-force period, if the 
pulling force is sufficiently high ($f_D \gg f_T$), the distribution relaxes to the stationary distribution in the linear potential well with a slope $f_D$
\begin{equation}
    p(x)= \frac{f_D e^{f_D (L+x)}}{e^{f_D L+f_D}-1}.
    \label{eq:stateDown}
\end{equation}

Inserting Eqs.~(\ref{eq:splitting}) and (\ref{eq:statState}) into Eq.~(\ref{eq:escapeProb}) one may obtain formula for the average number of force periods before escape
\begin{equation}
    \langle n \rangle = \frac{2 e^{f_T L} (f_T \sinh f_U-f_T \cosh f_U+f_U+f_T)-2 f_U}{f_U \left(e^{f_T L}-1\right)}.
    \label{eq:periodNumber}
\end{equation}

Another important time scale in the system is the mean time necessary to return to the origin from $x=1$ in the absence of the pulling force. Biologically, this corresponds to the protein refolding time. 
The latter can be calculated analogously to the left barrier problem considered above, using a formula for MFET from the interval limited by a reflecting boundary from one side \cite{gardiner2009}
\begin{equation}
\tau(x)= \int_{0}^{x} \frac{d y}{\psi(y)} \int_{y}^{1}  \psi(z) d z,
\label{eq:mfpt-ra}
\end{equation}
where
\begin{equation}
    \psi(x) = \exp{\left( \int_{0}^x f_U dy \right)}=\exp{\left(- f_U x \right)}.
    \label{eq:auxilaryfunction}
\end{equation}
Performing the integration in \eqref{eq:mfpt-ra} leads to
\begin{equation}
    \tau_\mathrm{fold} =\frac{f_U+e^{-f_U}-1}{f_U^2}
    \label{eq:returnTime}
\end{equation}
For large $f_U$ the folding proceeds deterministically, in a downhill manner, and $\tau_\mathrm{fold} \approx f_U^{-1}$.

In the limit of very short force periods, $t \ll \tau_\mathrm{fold} $, the translocation time can again be estimated analytically. This time, the starting point is the jammed configuration ($x=1$), where the system is trapped after the first force period. The probability that the particle, starting at $x=1$ can reach $x=-L$ over time $T$ can be estimated in the following manner. First, we note that in the limit of very short times, the deterministic drift of a particle under the action of $f_U$ and $f_T$ can be neglected with respect to diffusion, since the latter scales as $\sqrt{t}$ and the former as $t$. The diffusive current at $x=-L$ due to the source near a reflective wall at $x=1$ is given by
\begin{multline}
J(t) = - \frac{2}{\sqrt{2 \pi t}} \partial_x  \left( e^{-\frac{(x - 1)^2}{2t}} -  e^{-\frac{(x + 2L+1)^2}{2t}}  \right)_{x=-L} \\ = \frac{2^{3/2} (L+1)}{\sqrt{\pi t^3}} e^{-\frac{(L+1)^2}{2 t}}
\end{multline}
By integrating this over t, we obtain the escape probability over a single force period of the form
\begin{equation}
P(T)=\int_0^T J(t) dt  = 4 \  \text{Erfc}\left(\frac{L+1}{\sqrt{2 T}}\right)
\end{equation}
with the mean escape time proportional to the inverse of $P$, i.e. 
\begin{equation}
\langle \tau \rangle = \frac{T}{P} \sim 
 \frac{(L+1)\sqrt{\pi T}}{4 \sqrt{2}} {\mbox{\Large\( e 
\)}}^{\displaystyle \frac{(L+1)^2}{2 T}}
\end{equation}
where the asymptotic behaviour of the error function has been used. 

As we see, both long and short force periods result in very long translocation times; with $\langle \tau \rangle  \sim T$ for large $T$ and $\langle \tau \rangle  \sim \text{exp}(1/T)$ for short $T$. We thus expect that  between these two extremes there exists an optimal force period, $T_{min}$ for which the translocation is the fastest. One can anticipate that $T_{min}$ should be of the order of the folding time, to allow enough time for the system to reach the KL state and then attempt the barrier crossing. In the next section, we investigate the existence of the minimum numerically. 

\section{Numerical results}

The intermediate force periods, between the asymptotes considered in the previous section, do not lend themselves to analytical analysis and we need to resort to numerical methods. To obtain a result for arbitrary $T$, the Euler-Maruyama method \cite{higham2001algorithmic,mannella2002} of integration of the Eq~(\ref{eq:langevin}) has been used.
The trajectories were simulated with the time-step $\Delta t=10^{-5}$ until particle crosses barrier located in $L=0.25$ when the first exit time was registered.
MFET was subsequently obtained by averaging the first exit times over $N=10^5$ trajectories.

\begin{figure}[!h]
    \centering
    \includegraphics[width=0.95\columnwidth]{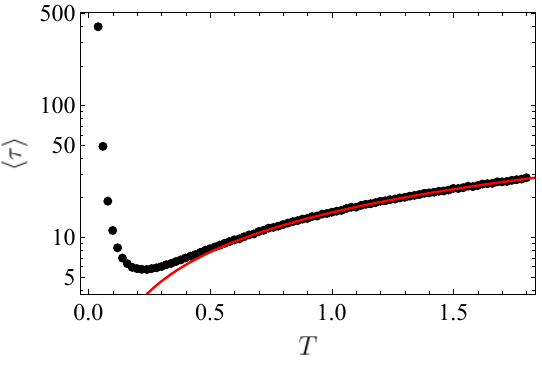}\\
    \includegraphics[width=0.95\columnwidth]{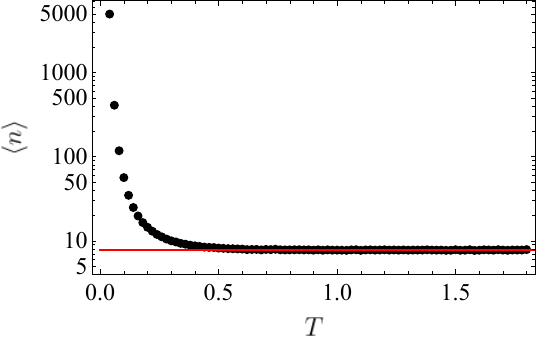}
    \caption{MFET (top panel) and the average number of force periods needed for a successful translocation (bottom panel) as a function of half-period $T$ for $f_T=1$, $f_U=1$, and $f_D=320$. Solid line corresponds to the asymptotic solution given by Eq~(\ref{eq:periodNumber}).
    }
    \label{fig:MFPTDown}
\end{figure}

\begin{figure}[!h]
    \centering
    \includegraphics[width=0.95\columnwidth]{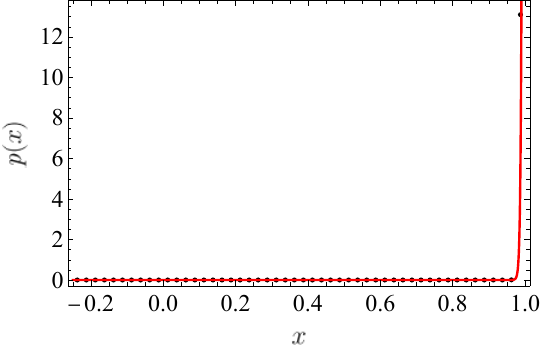}\\    
    \includegraphics[width=0.95\columnwidth]{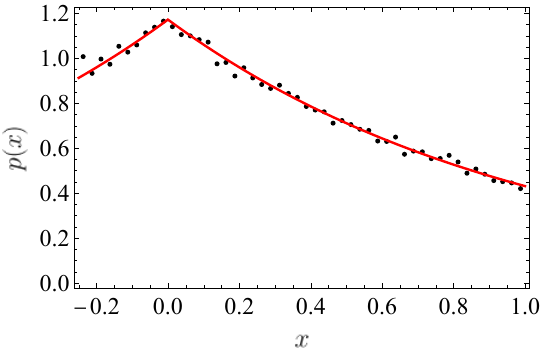}
    \caption{Probability densities for $T/2$, when the protein is pulled into the pore (top panel), and $5T/4$, when the force is switched off (bottom panel), for $f_T=1$, $f_U=1$, $f_D=320$ and $T=2$. Points correspond to the numerical results whereas solid lines represents approximated analytical solutions given by Eq.~(\ref{eq:stateDown}) (top panel) and  by Eq.~(\ref{eq:statState}) (bottom panel) respectively. 
    }
    \label{fig:pdfLong}
\end{figure}

%
%
We begin with an examination of the mean exit time (MFET) as a function of half-period $T$. Fig.~\ref{fig:MFPTDown} shows $\langle \tau(T) \rangle$ dependence for  $f_T=1$, $f_U=1$, and $f_D=320$. As expected, we see an eexponential increase of $\langle \tau \rangle$ for small periods and linear growth for large periods, with a minimum in between. 
For the example presented in Fig.~\ref{fig:MFPTDown} the minimum  corresponds to $\tau \approx 6$ and is located at $T_{min}\approx 0.2$. Note that this is slightly shorter than the protein refolding time, which, based on \eqref{eq:mfpt-ra} for this choice of parameters is $\tau_\mathrm{fold}  = e^{-1}\approx 0.37$. 
The linear growth of $\langle \tau(T) \rangle$ can be directly verified by inspection of $\langle n \rangle$. As can be seen in the bottom panel of Fig.~\ref{fig:MFPTDown}, the solution tends rapidly to the asymptotic behaviour given by Eq.~(\ref{eq:periodNumber}). The good agreement between the numerical data and the analytical solution for large force periods can also serve as a test of the approximations adopted when deriving Eq.~\eqref{eq:assimptitics}, in particular the assumption that the distribution can be approximated as stationary, Eq.~\eqref{eq:statState}. This is further confirmed by the analysis of the particle distributions, as shown in Fig.~\ref{fig:pdfLong}. 
As we see, in the first half of the period, the particles are pulled towards the KT state, relaxing to a delta-like distribution, due to the large depth of the energy well. On the other hand, in the second part of the period, the particles move towards the KL minimum, with the distribution relaxing towards that described by Eq.~(\ref{eq:statState}). 
\begin{figure}[!h]
    \centering
    \includegraphics[width=0.95\columnwidth]{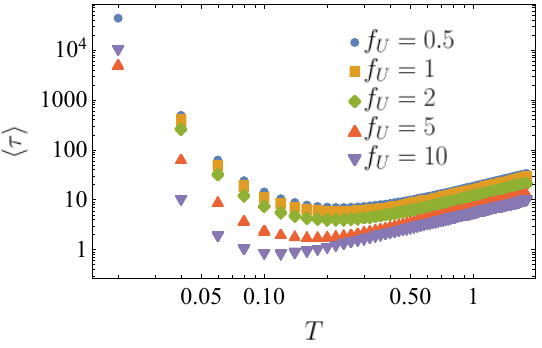}\\
    \includegraphics[width=0.95\columnwidth]{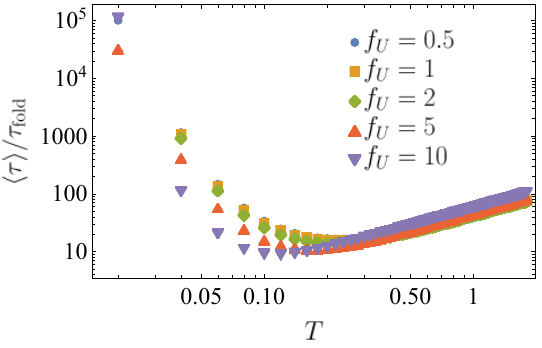}
    \caption{Mean first exit time as a function of the force period, $T$ for $f_T=1$, $f_D=320$ and a range of different $f_U$.
    }
    \label{fig:MFPTTrans3}
\end{figure}

The potential (\ref{eq:potential}) used in the model is described by three characteristic forces -- $f_T$ controlling the exit probability, $f_U$ connected to the folding and $f_D$ describing the action of the molecular motor, which - combined with steric interactions between the protein and the pore - tend to tighten the knot and block the pore. The exact value of $f_D$ does not affect the MFET, as long as it can be considered large, i.e., the system trapped by this force is basically unable to escape from $KT$ state. Weaker $f_D$ forces would be expected to facilitate escape from the system, since the particle would be able to reach the absorbing barrier at $x=-L$ even after venturing into the  region of potential well, i.e. $x>0$.

Contrary to $f_D$, increasing $f_U$ results not only in faster escape times but also in the emergence of a deeper minimum of MFET, shifted towards shorter force periods, as observed in the upper panel of Fig.~\ref{fig:MFPTTrans3}. This behavior can be explained by the shorter time required on average for the particle to return to the origin after being trapped in $KT$ state at $x=1$. Since this timescale is related to a refolding time,  $\tau_\mathrm{fold} $, we can expect that rescaling by it should account for the most of MFET dependence on $f_U$. This is indeed the case, as illustrated in  Fig.~\ref{fig:MFPTTrans3}, however for large $f_U$ the escape times become significantly shorter than expected by a simple rescaling by $\tau_\mathrm{fold} $. This can be rationalized by noting that - for the particle to have a chance of escaping - it needs to reach the left part of the potential $x<0$. The higher values of $f_U$ not only shorten the time necessary to reach this region but also increase the fraction of particles in the negative ($x<0$) part of the potential, relative to those in the positive ($x>0$) part during the off-force period.

\begin{figure}[!h]
    \centering
    \includegraphics[width=0.95\columnwidth]{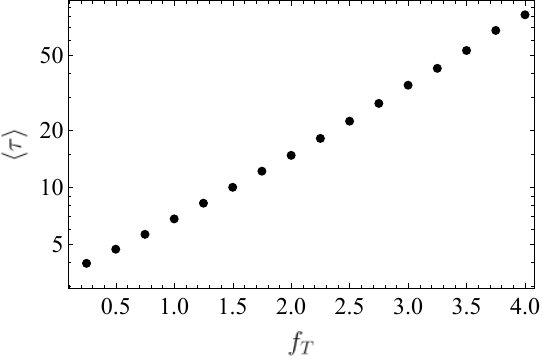}
    \caption{Mean first exit time as a function of $f_T$ for  $f_U=1$, $f_D=320$ and $T=0.2$. 
    }
    \label{fig:MFPTTrans}
\end{figure}

Finally, $f_T$ controls the probability of unknotting i.e. the larger $f_T$ the slower is the particle escape from the topological trap. To the leading order, the dependence of MFET on $f_T$ is given by the Arrhenius law, thus even a small increase of $f_T$ results in an exponential growth of MFET. In the asymptotic regime, $T\to \infty$, this behavior is predicted by Eq.~(\ref{eq:periodNumber}), however it holds in a much wider range of force period values. This is confirmed by Fig.~\ref{fig:MFPTTrans}, which shows the dependence of MFET on the transition force $f_T$ for $f_U=1$, $f_D=320$ and $T=T_{min}=0.2$ (the latter corresponds to the minimum MFET as a function of $T$ for $f_U=f_T=1$ and $f_D=320$).     
Despite the fact that $T$ is relatively small, definitely not in the $T \rightarrow \infty$ asymptotic regime, one can still observe the linear behavior in the logarithmic plot.  

%
%
\section{Summary and conclusions \label{sec:summary}}

We have constructed a simple energy landscape model of the protein translocation process, with a time-periodic potential, mimicking the cyclic nature of biological motors. The modal is solvable both in the limit of very short and very long periods of the driving force. In both of these limits the translocation time diverges, which suggests that there is an optimum force period corresponding to the shortest translocation time. We find this time numerically and show that it is of the order of the protein refolding time, $\tau_\mathrm{fold}$ (for relatively low folding forces) or shorter than $\tau_\mathrm{fold}$ (for large folding forces resulting in folding times shorter than the force period). Importantly, the dwell times between the power strokes of the biological motors are distributed rather broadly, with two characteristic timescales; one of the order of hundreds of milliseconds and the other of about ten seconds \cite{rief2000,Kotamarthi2020}. On the other hand, protein folding times are also broadly distributed~\cite{Naganathan2005}, with smaller two-state proteins folding on a millisecond scale, while more complex molecules folding on the scale of seconds. Similar timescales of motor action and protein folding mean that translocases act near the optimum. 

These results are in agreement with molecular dynamics results of Ref.~\cite{szymczak2016}, but the stochastic model has the benefit of being simple and easily interpretable in terms of a handful of parameters only (folding time, motor force intensity and frequency), making it easier to understand the key factors controlling system dynamics. 

Finally, we note that, in principle, an analogous model can also be used to simulate the escape of translocating proteins from other kinetic traps, not necessarily of a topological nature. One such example was reported in Ref.~\cite{Tian2005}, which shows that during the translocation of barnase through a mitochondrial pore, the protein can get trapped in a long-lived intermediate state, which blocks the pore and stalls the translocation. Again, the repetitive forces can lead the system out of such a kinetic trap, and it seems that the model presented here can be applied if one interprets KT as an unfolding intermediate which acts as a  kinetic trap and KL as a native state of this protein.

%
%
\section*{Acknowledgements}

This research was supported in part by PLGrid Infrastructure. We also acknowledge the support of Descartes project 
(POWR.03.02.00–00-I026/17-00) co-financed by the European Union through the European Social Fund under the Operational Programme Knowledge Education Development.

\section*{Data availability}
The data that support the findings of this study are available from one of the authors (KC) upon reasonable request.

%
%

%
%


%

\end{document}